\documentclass[12pt,a4paper]{article}
\usepackage{type1cm,amsmath,hangcaption,graphicx,indentfirst}
\usepackage[psamsfonts]{amssymb}
\numberwithin{equation}{section}

\begin{document}
\begin{titlepage}

 \renewcommand{\thefootnote}{\fnsymbol{footnote}}
\begin{flushright}
 \begin{tabular}{l}
 KEK-TH-1043\\
 UT-05-15 \\
 hep-th/0510129\\
 \end{tabular}
\end{flushright}

 \vfill
 \begin{center}
 \font\titlerm=cmr10 scaled\magstep4
 \font\titlei=cmmi10 scaled\magstep4
 \font\titleis=cmmi7 scaled\magstep4
 \centerline{\titlerm D-instantons and Closed String Tachyons}
 \vskip .3 truecm
 \centerline{\titlerm in Misner Space}
 \vskip 2.5 truecm

\noindent{ \large Yasuaki Hikida$^a$\footnote{E-mail:
hikida@post.kek.jp} and Ta-Sheng Tai$^b$\footnote{E-mail:
tasheng@hep-th.phys.s.u-tokyo.ac.jp}}
\bigskip

 \vskip .6 truecm
\centerline{\it $^a$ Theory Group, High Energy Accelerator Research Organization (KEK)}
\centerline{\it Tukuba, Ibaraki 305-0801, Japan}
\bigskip
\centerline{\it $^b$ Department of Physics,  Faculty of Science, 
University of Tokyo}
\centerline{\it Hongo, Bunkyo-ku, Tokyo 113-0033, Japan}

 \vskip .4 truecm

 \end{center}

 \vfill
\vskip 0.5 truecm

\begin{abstract}

We investigate
closed string tachyon condensation
in Misner space, a toy model for 
big bang universe.
In Misner space, we are able to condense tachyonic 
modes of closed strings in the twisted sectors,
which is supposed to remove the big bang singularity.
In order to examine this, we utilize D-instanton as a probe. 
First, we study general properties of D-instanton
by constructing boundary state and 
effective action.
Then, resorting to these, we are able to show that tachyon condensation actually deforms the geometry
such that the singularity becomes milder.

\end{abstract}
\vfill
\vskip 0.5 truecm

\setcounter{footnote}{0}
\renewcommand{\thefootnote}{\arabic{footnote}}
\end{titlepage}

\newpage

\tableofcontents
\section{Introduction}
\label{Intoduction}

In general relativity, we may encounter singularities at
the early universe or inside black holes, and stringy effects
are supposed to resolve these singularities.
However, it is difficult to show this explicitly
since string theory on these backgrounds is not solvable
in general.
Therefore, it is useful to investigate simpler models for 
cosmological backgrounds by utilizing Lorentzian orbifolds as in
\cite{KOSST,BHKN,CC,Nekrasov,Simon,LMS1,CCK,Lawrence,LMS2,FM,HP}
or gauged WZW models as in \cite{NW,EGKR,CKR,HT,TT}.
In this paper, we consider a simplest model for the early
universe, namely Misner (Milne) space \cite{Misner}.
Strings in this background are investigated in 
\cite{Nekrasov,BCKR,Pioline1,Pioline2,Pioline3,Pioline4}
and D-branes are in \cite{HNP2}.%
\footnote{See, e.g., \cite{HS,AP,DN,CLO,Okuyama,HNP} for D-branes in 
other time-dependent backgrounds.}

Misner space can be constructed as (1+1) dimensional
Minkowski space with the identification of a discrete boost.
This space consists of four regions, which are classified into
two types. One type are called as cosmological regions,
which include big bang/big crunch singularity, and 
the other are called as whisker regions, 
which include closed time-like curves.
In bosonic string or superstring theories with opposite
periodic condition of space-time fermions,  
there are tachyonic modes
in the twisted closed string sector. We find that
there are two types of tachyonic modes, and one of them
is localized in the whisker regions.

It is therefore natural to expect that the condensation
of the tachyon removes the regions with closed time-like curves \cite{Hagedorn}. 
Furthermore, it still remains crucial 
if the condensation affects the cosmological regions 
through the big bang singularity 
and ultimately resolves the singularity \cite{MS} 
(see also \cite{Silverstein}).
This is an analogous situation to the condensation of
localized tachyon in Euclidean orbifolds, say, 
${\mathbb C}/{\mathbb Z}_N$ \cite{APS}.
See also \cite{Martinec,HMT} and references therein.

In order to investigate the tachyon condensation,
we utilize D-instantons in the Misner space.%
\footnote{While completing this work, a paper analyzing 
similar situation appeared in the arXiv \cite{She},
where the analysis in null brane case \cite{BKRS} was 
directly applied.
The relation to matrix models has been discussed 
recently in \cite{CSV,Li,LS,pp,Chen,RS} as well.}
We find that there are D-instantons localized in the
big crunch/big bang singularity, which are analogous
to fractional branes in orbifold models \cite{DDG,DG,BCR}.
Due to this analogy, we call this type of instantons as 
``fractional'' instantons.
By summing every fractional instantons, we can construct
a type of D-instanton away from the big crunch/big bang
singularity. We construct effective theory for open strings
on this instanton, and probe the background geometry 
using its moduli space.
Before condensing the tachyon, 
the geometry read off from the moduli is Misner space
as expected. After turning on a small tachyon vev,
we find the singularity becomes milder than the original
background, which is consistent with the conclusion in \cite{MS}.
Concretely, we find a flow from Misner space with smaller space cycle
to larger space cycle and a flow from Misner space to the future patch
of Minkowski space in Rindler coordinates.

The organization of this paper is as follows.
First, we introduce Misner space as a Lorentzian orbifold,
and study closed strings both in untwisted and twisted
sectors. In particular, we closely investigate the
properties of the tachyonic modes of closed strings
in twisted sectors, and observe one type of the tachyon
is localized in the whisker regions.
In order to see how the condensation of the tachyonic modes
changes the background geometry, we utilize D-instantons
since they are suitable to probe the deformed geometry \cite{APS,BKRS}.
Using the boundary state formalism, we confirm the existence
of fractional instantons.
We construct effective theory for open strings on the sum
of every fractional instantons, and probe the geometry following 
\cite{APS}.
{}From this probe, we see that the deformation by the condensation
of the closed string tachyon tends to make the big bang singularity milder.

\section{Closed string tachyons in Misner space}
\label{tachyon}

For string theory application, it is convenient to
define Misner space \cite{Misner} as
(1+1) dimensional Minkowski space with
identification through Lorentz boost.
We denote the light-cone coordinates as
$x^{\pm} = \frac{1}{\sqrt2} (x^0 \pm x^1)$,
then the Misner space is defined by using the identification
\begin{align}
x^{\pm} &\sim g \cdot x^{\pm} ~,
& g \cdot x^{\pm} &= e^{\pm 2 \pi \gamma} x^{\pm} ~,
\label{misner}
\end{align}
where  $\gamma \in {\mathbb R}$.

The space is divided into four regions by lines
$x^+ x^- = 0$.
We are mainly interested in $x^+ , x^- > 0$ region,
which can be regarded as a big bang universe.
In this region, we change the coordinate by
$x^{\pm} = \frac{1}{\sqrt{2}}t e^{\pm \psi}$,
then the metric is
\begin{align}
ds^2 = -dt^2 + t^2 d \psi ^2
\label{cosmologicalr}
\end{align}
with the periodicity $\psi \sim \psi + 2 \pi \gamma$.
The radius of the space cycle depends on the time $t$,
and the radius vanishes when $t=0$. Thus, the space
starts at the big bang and expands as time goes.
The region with $x^+ , x^- < 0$ is the time reversal
of the big bang universe, namely a big crunch universe.
The metric is also given by \eqref{cosmologicalr}.
The space starts at $t=-\infty$ with infinitely large
radius, shrinks as time goes, and meets the big
crunch when $t=0$. We can deal with this region
in a way similar to the big bang region by
time reversal.

The regions with $x^+ x^- > 0$ are called as 
cosmological regions, and the other two
regions with  $x^+ x^- < 0$ are referred to
whisker regions.
Under the coordinate transformation
$x^{\pm} = \pm  \frac{1}{\sqrt{2}} r e^{\pm \chi}$,
the metric in the whisker regions is given by
\begin{align}
ds^2 = d r^2 - r^2 d \chi^2 ~.
\label{whiskerr}
\end{align}
There are closed time-like curves everywhere in these regions,
which are attributed to 
the periodicity $\chi \sim \chi + 2 \pi \gamma$ of time coordinate. 
We will not study the whisker regions in detail,
but it was suggested in $\cite{Hagedorn}$ that these regions might be removed from the space-time
by tachyon condensation.

In addition to Misner space, $d$ extra flat directions are included to form critical string theory, 
where $d=24$ for bosonic case and $d=8$ for superstring case. In this section, 
the untwisted sector and the twisted sectors for bosonic case are analyzed at first, and superstring cases follow soon.
Investigation of the behavior of tachyonic modes is of main interest,%
\footnote{Similar discussion was made in \cite{Hagedorn}
for tachyonic modes in O-plane.}
whose condensation will be considered below.

\subsection{Closed strings in the untwisted sector}

Let us first examine the untwisted sector of closed strings.
To have a quick grasp of the behavior of the closed strings, 
it is sufficient to focus on the zero mode part, nevertheless higher modes can be included easily.
The untwisted sector is obtained by summing all images
of the discrete boost \eqref{misner} in the covering space,
that is, two dimensional Minkowski space-time.
The zero mode part reduces to  a particle in the untwisted sector,
and its trajectory in the covering space is
a straight line
\begin{align}
 X^{\pm} = x_0^{\pm} + \alpha ' p^{\pm} \tau ~.
\end{align}
The mass square is given by $m^2 = 2 p^+ p^-$, which can be 
positive or negative in the bosonic string or type 0 superstring 
theory. For massive modes $p^+, p^- > 0$,\footnote{The particle
with $p^+, p^- < 0$ travels from future to past, or it can
be considered as an anti-particle.}
the particle starts from infinite past in the big crunch region
and ends at infinite future in the big bang region.
While cruising over the space-time, the particle crosses a whisker
region only up to a point $| r_0 | = |l|/m$ with 
$l=x^+_0 p^- - x^-_0 p^+$.
For the tachyonic modes, we have to set $p^+ < 0,p^- > 0$
or  $p^+ > 0,p^- < 0$.
The trajectory begins from spatial infinity in a whisker region, 
crosses a cosmological region and terminates at the spatial 
infinity in the other whisker region. 
In other words, the tachyonic modes exist
only for a short time $0 < t < t_0=|l|/\sqrt{- m^2}$ 
or $0 > t > t_0= -|l|/\sqrt{- m^2}$ in 
cosmological regions.

When it comes to the bosonic string case, the Virasoro condition reads
\begin{align}
L_0 + \tilde L_0 -2 & = \frac{\alpha '}{2} ( - 2 p^+ p^- + m^2)  = 0 ~,
&m^2 &= \vec k ^2 + \frac{2}{\alpha '} (N + \tilde N - 2) ~,
\end{align}
where $\vec k$ represents the momenta along the extra
directions and $N,\tilde N$ denote occupation numbers.
For our analysis, it is convenient to rewrite as
\begin{align}
 p_{\eta}^2  + V(\eta) &=  0~,
&p_{\phi}^2 + V(\phi) &= 0 ~,
\label{dis}
\end{align}
where
\begin{align}
V(\eta) &= - l^2 - m^2 e^{2 \eta}  ~,
&V(\phi) &= - l^2 + m^2 e^{2 \phi} ~.
\label{potential}
\end{align}
We have changed the coordinates as $t = \pm e^{\eta},r = \pm e^{\phi}$.
For massive modes $m^2 > 0$, the condition $p^2_{\eta} \geq 0$ is 
always satisfied, thus the string runs from infinite past to 
infinite future. 
On the other hand, the condition $p^2_{\phi} \geq 0$ restricts
$|r| \leq |l|/m$.
For tachyonic modes $m^2 < 0$, the condition $p^2_{\phi} \geq 0$
is always satisfied, while the condition $p^2_{\eta} \geq 0$
restricts $|t| \leq |l|/\sqrt{- m^2}$.
These results are consistent with the previous analysis.

For a scalar particle, Klein-Gordon equation can be
obtained by replacing $p_{\eta},p_{\phi}$
with derivatives $i\partial_{\eta},i\partial_{\phi}$.
The wave function shows oscillatory or damping behavior
above or below the potential.
This implies that the wave function is localized in the cosmological
regions for massive modes and in the whisker regions for
tachyon modes. In fact, the wave function can be expressed 
in the covering space as \cite{Nekrasov}
\begin{align}
\Psi_{p,l} = \int d w e^{i(p^+ X^- e^{-\gamma w}+ p^- X^+ e^{\gamma w} + i w l + \vec k \cdot \vec X)} ~,
\end{align}
which is invariant under the orbifold action.
For massive modes, we can set $p^+ = p^- = m/\sqrt2 $,
then the wave function is localized in the cosmological regions.
In particular, for large $ - X^+X^- \gg |l|/m$, the wave function 
shows damping behavior \cite{Nekrasov} as 
$\Psi_{p,l} \propto \exp (- m r)$.
For tachyonic case, we may set 
$p^+ = - p^- = \sqrt{- m^2 /2 }$ or
$p^+ = - p^- = -\sqrt{- m^2 /2 } $.
Then, the wave function is localized in the whisker regions
and shows damping behavior in the cosmological regions as
$\Psi_{p,l} \propto \exp (- \sqrt{- m^2} |t|)$ for 
$X^+ X^- \gg |l|/\sqrt{-m^2}$.

\subsection{Closed strings in the twisted sectors}

In the orbifold theory, there are twisted sectors
of closed strings. Because of the identification \eqref{misner}, twisted periodic conditions give rise to
\begin{align}
 X^{\pm} (\tau, \sigma+ 2\pi) = e^{\pm 2\pi
\gamma w} X^{\pm}(\tau, \sigma)
\label{twist}
\end{align}
with non-trivial $w \in {\mathbb Z}$. 
There are two types of twisted closed strings, 
since two types of cycles, on which closed strings
can be wrapped, namely, $\psi$-cycle in the cosmological
regions and $\chi$-cycle in the whisker regions.

The lowest mode of the
solution to \eqref{twist} is
\begin{align}
X^{\pm} (\tau, \sigma) &= \pm \sqrt{\frac{\alpha '}{2}} \frac{\alpha^{\pm}_0}{\nu} e^{\pm \nu (\tau + \sigma) }
 \mp \sqrt{\frac{\alpha '}{2}} \frac{\tilde \alpha^{\pm}_0}{\nu} e^{\mp \nu (\tau - \sigma) } ~,
 & \nu &= \gamma w 
 \label{modes}
\end{align}
which would provide clear physical picture of the winding strings.
For bosonic strings, the Virasoro constraints read
\begin{align}
\omega^2 &= \frac{1}{2}(\alpha^+_0 \alpha^-_0 + \alpha^-_0 \alpha^+_0) ~,
& \tilde \omega^2
= \frac12( \tilde \alpha^+_0 \tilde \alpha^-_0 + \tilde \alpha^-_0\tilde
\alpha^+_0)
\end{align}
with
\begin{align}
\omega^2 &= \frac{\alpha '}{4} \vec k ^2 + N - 1 + \frac12 \nu^2 ~,
&\tilde \omega^2 &= \frac{\alpha '}{4} \vec k ^2 + \tilde N - 1 + \frac12 \nu^2 ~.
\label{tomega}
\end{align}
{}From the above equations, we can see that 
$\omega ^2 , \tilde \omega ^2$ can take positive and negative
values for small $ | \nu | < \sqrt2$.
As shown below, there is no shift of zero point energy 
$\frac12 \nu^2$ in superstring theories.
Under the lowest mode expansion,
the level matching condition demands
$\omega ^2 = \tilde \omega ^2$, 
thus there is no angular momentum 
$\omega ^2 - \tilde \omega ^2 = \nu l=0$.
In fact, it was shown in \cite{HNP2} that
the general on-shell states are
isomorphic to the states with only lowest modes
in Misner space part.

\begin{figure}
\centerline{\scalebox{0.6}{\includegraphics{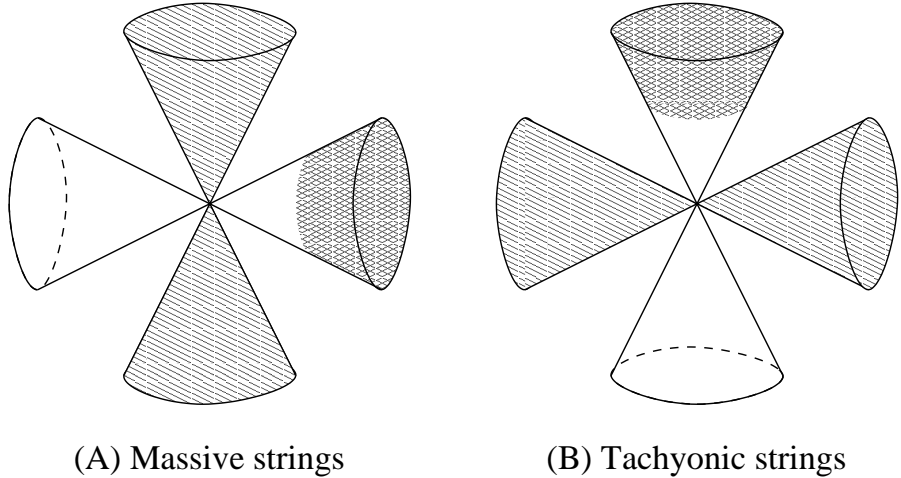}}}
\caption{\it Winding strings with no angular momentum $l=0$.
(A) For massive case,
winding strings are wrapped on the whole cosmological regions
or a part of whisker regions from finite $r_0$ to spatial infinity. 
(B) For tachyonic case,
winding strings are wrapped on the whole whisker regions
or a part of cosmological regions from finite $t_0$ to past or future 
infinity.}
\label{winding}
\end{figure}

For this reason, we will only consider the case with $l=0$ in the following.
For massive case $\omega^2 >0$,
we can set \cite{Pioline1}
\begin{align}
 \alpha^+_0 &= \alpha^-_0 = \epsilon \omega  ~,
 &\tilde \alpha^+_0 &= \tilde \alpha^-_0 = \tilde \epsilon \omega  ~,
 &\epsilon,\tilde \epsilon &= \pm 1
\end{align}
without loss of generality.
With these solutions, we obtain
\begin{align}
X^{\pm} (\tau,\sigma) =
 \frac{\sqrt{2 \alpha '}\omega}{\nu} e^{\pm \nu \sigma} \sinh \nu \tau
 \label{short}
\end{align}
for $\epsilon = \tilde \epsilon = 1$.
This represents a string wrapping the whole cosmological regions,
namely, winding $\psi$-cycle.
Another choice may be
$\epsilon = - \tilde \epsilon = 1$, which leads to
\begin{align}
X^{\pm} (\tau,\sigma ) =
\pm \frac{\sqrt{2 \alpha '}\omega}{\nu} e^{\pm \nu \sigma} \cosh \nu \tau ~.
 \label{long}
\end{align}
This string winds $\chi$-cycle in the whisker regions,
and exists from $r_0= 2 \sqrt{\alpha '} \omega / \nu$ to spatial infinity.
See fig. \ref{winding} for these classical trajectories.

For tachyonic case $\omega^2 =  - {\cal M}^2 < 0$,
we can set 
\begin{align}
 \alpha^+_0 &= - \alpha^-_0 = \epsilon {\cal M}  ~,
 &\tilde \alpha^+_0 &= - \tilde \alpha^-_0 = \tilde \epsilon {\cal M}  ~,
 &\epsilon,\tilde \epsilon &= \pm 1 ~.
\end{align}
When $\epsilon = \tilde \epsilon = 1$,
the classical trajectory is 
\begin{align}
X^{\pm} (\tau,\sigma) = \pm
 \frac{\sqrt{2\alpha'} {\cal M}}{\nu} e^{\pm \nu \sigma} \sinh \nu \tau ~,
\end{align}
which wraps on the whole whisker region.
A good property for us is that the tachyon touches the cosmological
regions only through the big crunch/big bang singularity.
This is analogous to the tachyons in twisted sectors of Euclidean 
orbifolds, which are localized at their fixed points.
Using this fact, the authors of \cite{APS} can discuss closed
tachyon condensation, since the condensation affects only
small regions of backgrounds.
Therefore, it is natural to think that we can also
analyze the closed tachyon condensation in the cosmological regions, 
even though it seems difficult to know the fate of the whisker 
regions.%
\footnote{It was suggested in \cite{Hagedorn} that the tachyon 
condensation removes the whisker regions from the background.}
When $\epsilon = - \tilde \epsilon = 1$,
the classical trajectory is 
\begin{align}
X^{\pm} (\tau,\sigma ) =
\frac{\sqrt{2\alpha'} {\cal M}}{\nu} e^{\pm \nu \sigma} \cosh \nu \tau ~,
\end{align}
which stands for the process of string pair creation
at a time $t_0 = 2 \sqrt{\alpha '} {\cal M}/\nu $ 
and ends at future infinity.%
\footnote{We assumed ${\cal M}/\nu > 0$. If ${\cal M}/\nu < 0$, 
then the trajectory represents the time-reversal process.}
We will not consider the condensation of this tachyon
because it has nothing to do with the resolution of big crunch/big bang
singularity.

In quantum level, the modes are treated as operators
with commutation relations
\begin{align}
 [ \alpha^+_0 , \alpha^-_0 ] &= - i \nu ~,
&[ \tilde \alpha^+_0 , \tilde \alpha^-_0 ] &= i \nu ~.
\end{align}
Following \cite{Pioline1}, we use a representation
\begin{align}
 \alpha ^{\pm}_0 &= i \sqrt{\frac{\alpha '}{2}} \partial_{\mp} \pm 
      \frac{\nu} {\sqrt{2 \alpha '}}  x^{\pm} ~,
 &\tilde \alpha ^{\pm}_0 &= 
 i \sqrt{\frac{\alpha '}{2}} \partial_{\mp} \mp 
  \frac{\nu}{\sqrt{2 \alpha '}} x^{\pm} ~.
\end{align}
Then, the Virasoro conditions \eqref{tomega} reduce
to the equations \eqref{dis}
with the potentials \cite{Pioline1}
\begin{align}
V(\eta) &= - \frac{\nu ^2 e^{4 \eta} }{ {\alpha '}^2} 
           - \frac{4 \omega ^2}{\alpha '} e^{2 \eta}  ~,
&V(\phi) &= -\frac{ \nu ^2 e^{4 \phi}}{ {\alpha '}^2}  
           + \frac{4 \omega^2}{\alpha '} e^{2 \phi} ~.
\end{align}
For massive case  $\omega ^2 > 0$, $p ^2 _{\eta} \geq 0$ does not
give any condition, but $p^2_{\phi} \geq 0$ is equivalent to
$| r | \geq 2 \sqrt{\alpha '} | \omega/\nu | $.
For tachyonic case  $\omega ^2 = - {\cal M}^2 < 0$,
$p^2_{\phi} \geq 0$ does not give any condition, but 
$p ^2 _{\eta} \geq 0$ is equivalent to 
$|t| \geq 2 \sqrt{\alpha '} | {\cal M}/\nu |$.
These conditions are the same as in the previous analysis.
Replacing $p_{\eta},p_{\phi}$ with derivatives 
$i \partial_{\eta}, i \partial_{\phi}$,
the Virasoro conditions lead to Klein-Gordon equation.
The solution should become oscillatory above the potential
and damping below the potential as mentioned before.
In each of the regions the wave functions can
be written in terms of Whittaker functions \cite{GS,Pioline1},
and in the whole space wave functions can be 
obtained by connecting those in the each region.

\subsection{Closed superstrings}

In this subsection, tachyonic modes 
in superstring cases are brought about 
by introducing worldsheet fermions and taking a particular GSO projections.
Due to the boundary condition \eqref{twist},
the full mode expansion for $X^{\pm}(\tau,\sigma)$ 
in $w$-th twisted sector is of the form
\begin{align}
 X^{\pm} (\tau,\sigma) = i \sqrt{\frac{\alpha '}{2}} 
  \sum_{n \in {\mathbb Z}} \left[
  \frac{\alpha^{\pm}_n}{n \pm i\nu} e^{-i(n \pm i\nu)(\tau + \sigma)} +
  \frac{\tilde \alpha^{\pm}_n}{n \mp i\nu} e^{-i(n \mp i\nu)(\tau - \sigma)}
  \right] ~,
\end{align}
where the oscillators satisfy the following commutation relations
\begin{align}
 [\alpha ^+_m,\alpha ^-_n] &= (- m - i \nu )\delta_{m+n} ~,
&[\tilde \alpha ^+_m, \tilde \alpha ^-_n] &= (- m + i \nu )\delta_{m+n} ~.
\end{align}
The boundary conditions for worldsheet fermions are
\begin{align}
 \psi^{\pm} (\tau , \sigma + 2 \pi)
&= - (-1)^A e^{\pm 2 \pi \nu} \psi^{\pm} (\tau , \sigma) ~,
&\tilde \psi^{\pm} (\tau , \sigma + 2 \pi)
&= - (-1)^A e^{\pm 2 \pi \nu} \tilde \psi^{\pm} (\tau , \sigma) ~,
\label{NSbc}
\end{align}
where $A=0$ for NS-sector and $A=1$ for R-sector.
The mode expansions can be given then
\begin{align}
 \psi^{\pm} &= \sum_{r \in  {\mathbb Z} + (A+1)/2}
  b^{\pm}_{-r} e^{-i(r \pm i \nu)(\tau + \sigma)} ~,
 &\tilde \psi^{\pm} &= \sum_{r \in  {\mathbb Z} + (A+1)/2}
  \tilde b^{\pm}_{-r} e^{-i(r \mp i \nu)(\tau - \sigma)} ~,
\end{align}
where the oscillators satisfy the anti-commutation relations
\begin{align}
 \{ b^{\pm}_r , b^{\mp}_s \} &= - \delta_{r+s} ~,
 &\{ \tilde b^{\pm}_r , \tilde b^{\mp}_s \} &= - \delta_{r+s} ~.
\end{align}

Combined with the bosonic part, the Virasoro constraint leads to%
\footnote{The counter part of right-mover is given in the same way.}
\begin{align}
\omega ^2 = \frac{\alpha '}{4} \vec k ^2 
 + N^{\pm}_{b} + N^{\pm}_{f} + N_{b} + N_{f} - \frac{1-A}{2} ~,
 \label{fvirasoro}
\end{align}
where $N_b,N_f$ are the occupation numbers of bosonic and
fermionic modes in extra $8$ directions.
Moreover,  $N^{\pm}_b,N^{\pm}_f$ are the occupation numbers of 
bosonic and fermionic modes of Misner space part.
They are explicitly given as
\begin{align}
 N^{\pm}_b = - \sum_{n > 0} \alpha^+_{- n} \alpha^-_{n}
             - \sum_{n > 0} \alpha^-_{- n} \alpha^+_{n}
\end{align}
for bosonic part and
\begin{align}
N^\pm_{f} = c(A) - \sum_{r > 0} (r - i \nu) b^+_{-r} b^-_r
   - \sum_{r > 0} (r + i \nu) b^-_{-r} b^+_r 
\end{align}
with $c(0) = 0$ and $c(1)= \frac{i \nu}{2} [ b_0^+ , b_0^- ]$
for fermionic part.
Notice that there is no zero energy shift in \eqref{fvirasoro}.

There appears to be tachyonic mode in NSNS sector, 
though it is projected out by the usual GSO projection. 
Nevertheless, it is possible to consistently impose 
usual GSO projection to even $w$-th sectors and
opposite GSO projection to odd $w$-th sectors.
More explicitly, we can use
\begin{align}
\begin{aligned}
 P^{\text {GSO} }_{\text {NSNS}} &= \frac14 \left(1 + (-1)^w(-1)^F \right)
                    \left(1 + (-1)^w(-1)^{\tilde F} \right) ~, \\
 P^{\text {GSO} }_{\text {RR}} &= \frac14 \left(1 + (-1)^w(-1)^F \right)
                    \left(1 \mp (-1)^w(-1)^{\tilde F} \right) ~,
                    \label{opposite}
\end{aligned}
\end{align}
where $F,\tilde F$ are the fermion number operators for
left and right moving parts, respectively. The sign $-$
is for type IIA and $+$ is for type IIB. 
The GSO projection for NSR or RNS sectors is defined 
in a similar manner.
Though the bulk tachyon is thrown away by this GSO projection,
in odd $w$-th twisted sectors tachyon mode survives.

The partition function with this GSO projection can be
evaluated as \cite{Nekrasov,HNP2}\footnote{%
In order to compute partition functions or construct boundary
states, it is convenient to adopt a different way to construct
Hilbert space. Namely, we regard $\alpha^+_0,\alpha^-_0$
as creation and annihilation operators. For more detail,
see \cite{Pioline1,HNP2}.}
\begin{align}
 \int \frac{d^2 \tau}{16 \pi^2 \tau_2^2 \alpha '}
  \sum_{k,w}
  \frac{|\vartheta_3 (y|\tau) \vartheta_3 (0|\tau)^3
        - (-1)^w \vartheta_4 (y|\tau) \vartheta_4 (0|\tau)^3
        - (-1)^k \vartheta_2 (y|\tau) \vartheta_2 (0|\tau)^3
        |^2}{(4 \pi^2 \tau_2 \alpha ')^3
             |\vartheta_1 (y|\tau ) \eta (\tau )^9|^2}
\end{align}
with $y=i \gamma ( w \tau + k )$. 
The Hilbert spaces of all $w$-th twisted sector were summed over, 
and the projection operator $P = \sum_k g^k$ was inserted.
The factor $(-1)^k$ is needed to ensure the modular invariance,
and it means that space-time fermions should
give factor $-(-1)^k$ when going around $\tau_2$-cycle.
In other words, the GSO projection \eqref{opposite}
is equivalent to the requirement that space-time fermions
should transform under the orbifold action, such that, $g^k \cdot S^{\pm} = (-1)^k e^{\pm k \pi \gamma} S^{\pm}$.
Note that this action is of the same form as in \cite{APS}.


\section{D-instanton probe of closed tachyon condensation}

As shown in the previous section, there are tachyonic modes in the twisted sectors in bosonic string or superstring theories
with our choice of GSO projection.
We will focus on the big bang region, where the 
tachyonic modes are localized at the big bang singularity.
It is expected that the condensation of the tachyon
alters the geometry near the big bang singularity
just like the tip of cone of Euclidean orbifold is
deformed by localized closed string tachyon \cite{APS}.

In order to examine the deformed geometry,
D-instantons are shown to be useful as probes
for their promised properties.
We will see that a sum of ``fractional'' instantons enable us 
to look into what happens after the tachyon condensation. 
By means of the effective action on
it, we observe that the geometry is deformed
and manage to answer whether the tachyon
condensation resolves the singularity.

\subsection{Open strings on instantons and boundary states}
\label{bs}

One way to construct instantons in Misner space is
to sum over all image instantons in the covering space.
Namely, if we put an instanton at 
$x^{\pm} = x^{\pm}_0$, then we have to sum
up instantons at 
$x^{\pm} = e^{\pm 2 \pi \gamma k} x^{\pm}_0$
for all $k \in {\mathbb Z}$.
We will call this type as ``bulk'' instanton.
In the covering space, these image instantons are
the same as in the flat space case, thus the
annulus amplitude for open strings between
image instantons and boundary states for each
image instantons are the same as in the flat space case.

Interestingly, we can build another type of instantons
invariant under the orbifold action.
They are localized at the big crunch/big bang singularity
$x^{\pm} = 0$, and analogous to so-called fractional branes
in Euclidean orbifolds \cite{DDG,DG,BCR}.
We will call this type as ``fractional'' instanton even though
they are not fractional with respect to bulk instanton in any sense.
Let us start from the bulk instanton at the singularity,
which is the sum of all $i$-th image instantons in the covering space.
The orbifold action $g$ acts both on Chan-Paton 
indices and the states as \cite{DM}
\begin{align}
 g :| \psi , i , j \rangle 
 \rightarrow \sum_{i',j'= - \infty}^{\infty}
 \gamma_{i,i'} | \hat g \psi , i' , j' \rangle \gamma_{j' , j}^{-1} ~.
 \label{CP0}
\end{align}
We set $\gamma_{i,i '} = \delta_{i+1,i '}$ because
the orbifold action shifts $i$-th image brane into
$(i-1)$-th image brane.
The orbifold action to the states is represented by $\hat g$.

Another basis for Chan-Paton indices, that fits more into our practice, is%
\footnote{We neglect an overall factor $1/N$ with $N=\sum_i 1$.}
\begin{align}
 | \psi , m , n \rangle  
 = \sum_{i,j= - \infty}^{\infty} 
\xi^{im}| \psi , i , j \rangle \xi^{-jn}
\label{basis}
\end{align}
with a complex number $\xi$.
This is more adequate since Chan-Paton indices are invariant
under the orbifold action as
\begin{align}
 g :| \psi , m , n \rangle 
 \rightarrow 
 \xi^{m - n} | \hat g \psi , m , n \rangle  ~.
\label{CP}
\end{align}
Thanks to the invariance of this basis,
the instantons, between which open strings are stretched, are
invariant under the orbifold, and they in fact differ from the 
bulk instanton as will be explained.
If we are dealing with the orbifold ${\mathbb C}/{\mathbb Z}_N$,
then  $g^N = 1$ should be satisfied for all $(m,n)$,
which leads to $\xi = e^{2 \pi i /N}$ and 
$(m,n) \in {\mathbb Z}_N$.
This suggests that there are $N$ types of fractional branes,
which transform differently under the orbifold action. 
Since there is no such a periodicity condition in our case,
an arbitrary $\xi$ can be used.
Later we fix $\xi$ such that the effective theory includes
desired fields.

To extract physical spectrum of open strings with Chan-Paton
indices $(m,n)$, we project into invariant subspace 
under the orbifold action by
\begin{align}
 P &= \sum_{k=-\infty}^{\infty} g^k ~, 
   &g : X^{\pm} \to e^{\pm 2 \pi \gamma } X^{\pm} ~.
\end{align}
Since the open strings on instantons satisfy the Dirichlet
boundary condition
\begin{align}
 X^{\pm} (\tau , \sigma = 0 ) &=0 ~,
&X^{\pm} (\tau , \sigma = \pi) &=0 ~,
\end{align}
the mode expansion is give by
\begin{align}
 X^{\pm} (\tau , \sigma ) =  \sqrt{2 \alpha ' } 
 \sum_{n \neq 0} \frac{\alpha^{\pm}_n}{n} 
 e^{in \tau} \sin (n \sigma ) ~.
\end{align}
In terms of oscillators, the Virasoro generator and  the
twist operator are written as
\begin{align}
 L_0 &=
 - \sum_{n \geq 1} \alpha^+_{-n} \alpha^-_n
 - \sum_{n \geq 1} \alpha^-_{-n} \alpha^+_n ~,
 &g &= e^{2 \pi \gamma i \hat J } ~,
 &i\hat J  &=  - 
  \sum_{n > 0} 
 \left( \frac{\alpha^{+}_{-n} \alpha^{-}_{n}}{n} 
   - \frac{\alpha^{-}_{-n} \alpha^{+}_{n}}{n} \right) ~.
\end{align}
Gathering all of them,
the one-loop amplitude of the open strings can be computed as%
\footnote{We implicitly neglect $k=0$ sector since this part
is the same as the bulk brane case.}
\begin{align}
 Z_{mn} (it) = \sum_{k= -\infty}^{\infty}
  {\rm Tr}_{{\cal H}_{mn}} g^k e^{-2 \pi t (L_0 - \frac{1}{12})}
  = \sum_{k= -\infty}^{\infty}\xi ^{k(m-n)}
    \frac{2 i \sinh (\pi |k| \gamma )\eta (it)}{\vartheta_1 (i |k|\gamma |it) }
 ~.
  \label{parto}
\end{align}
Recall that 
the orbifold action also acts on the Chan-Paton indices 
as in \eqref{CP}.
Using the modular transformation $t \rightarrow s=1/t$,
the partition function may be expressed as
\begin{align}
 Z_{mn} (i s) =  \sum_{k= -\infty}^{\infty}\xi ^{k(m-n)}
    \frac{2 \sinh (\pi |k| \gamma )\eta (i s) e^{- \pi s k^2 \gamma ^2}
  }{\vartheta_1 ( |k|\gamma s|i s) }~,
 \label{part}
\end{align}
which can be regarded as scattering between boundary states 
as shown below.

The above expression can be re-derived by path integral method as well. 
We use Euclidean worldsheet $(\sigma_1 ,
\sigma_2 )$ with $0 \leq \sigma_1 \leq \pi$ and $\sigma_2 \sim
\sigma_2 + 2\pi$, where the worldsheet metric and the Laplacian
are given by
\begin{align}
 ds^2 &= d^2 \sigma_1 + t^2 d^2 \sigma_2 ~,
&\Delta &= \frac{1}{t^2}(\partial_2^2 + t^2 \partial^2_1) ~.
\end{align}
We assign Dirichlet condition at the boundaries of worldsheet
as
\begin{align}
 X^{\pm} (\sigma_1=0 , \sigma_2) &=0 ~,
&X^{\pm} (\sigma_1= \pi , \sigma_2) &=0 ~.
 \label{D1bcpi1}
\end{align}
Under the identification of discrete boost \eqref{misner},
$k$-th twisted sector along $\sigma_2$ obeys
\begin{align}
 X^{\pm} (\sigma_1 , \sigma_2 + 2 \pi) =
  e^{\pm 2 \pi \gamma k} X^{\pm} (\sigma_1 , \sigma_2) ~.
 \label{D1bcpi2}
\end{align}
Conditions \eqref{D1bcpi1} and \eqref{D1bcpi2} altogether are solved by
\begin{align}
 X^{\pm} (\sigma_1, \sigma_2) =
\sum_{m \in {\mathbb Z},n > 0} a^{\pm}_{m,n}
  e^{i (m \mp i k\gamma) \sigma_2}\sin (n\sigma_1) ~.
\end{align}
Assuming the open string has the Chan-Paton
indices $(m,n)$ as before, the prescription in \eqref{CP} gives an overall factor $\xi^{k(m-n)}$ to the $k$-th
twisted sector.
Therefore, the partition function is summarized as%
\footnote{See, e.g., \cite{HNP2} for the detailed calculation.}
\begin{align}
\begin{aligned}
Z_{mn} (it) &= \sum_{k = - \infty}^{\infty} \xi^{k(m-n)}  {\rm Det}^{-1} (- \Delta ) \\
 & = \sum_{k = - \infty}^{\infty} \xi^{k(m-n)} \left[ \prod_{m,n > 0} (m+itn-ik\gamma)(m-itn-ik\gamma) \right]^{-1} ~,
\end{aligned}
\end{align}
which reproduces \eqref{parto}.

Next, we construct boundary states, which show how closed
strings couple to instantons.
Since the winding strings have zero size at the big crunch/big
bang singularity, the fractional instantons can couple to these 
winding strings.
Boundary states in the $k$-th twisted sector have to satisfy
\begin{align}
(\alpha^\pm_{n} - \tilde{\alpha}^\pm_{-n}) 
| B_{x^+,x^-}, k \rangle\rangle = 0 
\label{inst_cond}
\end{align}
for all $n \in {\mathbb Z}$, which
implies that the instantons can couple to
winding strings wrapping $\psi$-cycle for massive
strings and $\chi$-cycle for tachyonic strings.
It is nice for us since we would like to condense the
tachyonic modes of this type.

The boundary state satisfying the condition 
$\eqref{inst_cond}$ is given by
\begin{align}
| B_{x^+,x^-}, k \rangle\rangle 
 = \exp \left(-\sum_{n\ge1}\frac{\alpha^+_{-n}\tilde{\alpha}^-_{-n}}{n + i\nu}
 - \sum_{n\ge0}\frac{\alpha^-_{-n}\tilde{\alpha}^+_{-n}}{n - i\nu} \right)
 |0\rangle ~,
 \label{ishibashi}
\end{align}
whose overlaps are computed as
\begin{align}
\langle\langle B_{x^+,x^-}, k |e^{-\pi{s}(L_0+\tilde{L}_0-\frac{1}{6})}| B_{x^+,x^-}, k ' \rangle\rangle 
          = \delta_{k,k'}
    \frac{\eta (is) e^{-\pi s k^2 \gamma ^2}}{\vartheta_1 (s | k| \gamma | is)} ~.
\end{align}
The boundary state for fractional instanton with label $n$
can be given in a linear combination of \eqref{ishibashi}.
The coefficients are determined by the fact that
the partition function \eqref{part} should be
written in terms of boundary states as
\begin{align}
 Z_{m,n} (i s) =
 \langle\langle m |e^{-\pi{s}(L_0+\tilde{L}_0-\frac{1}{6})}| n \rangle\rangle ~.
\end{align}
The results are
\begin{align}
  | n \rangle\rangle 
&= \sum_{k=-\infty}^{\infty} \xi^{-kn} {\cal N}_k  
     | B_{x^+,x^-}, k \rangle\rangle ~,
  &\langle \langle n |
&= \sum_{k=-\infty}^{\infty} \xi^{kn}  {\cal N}_k 
     \langle \langle B_{x^+,x^-}, k | 
\end{align}
with ${\cal N}_k = \sqrt{ 2 \sinh (  \pi |k| \gamma ) }$.
In order to deal with D-instantons in critical bosonic string
theory, the sector with $24$ free boson ought to be added.
The superstring generalization is demonstrated in appendix B.


\subsection{Effective action of open strings on instantons}
\label{matrix}

We are going to build low energy effective action of
open strings on the fractional instantons presented above.
As shown above, there are tachyonic modes in 
either bosonic strings or superstrings with opposite
GSO projection to odd twisted sectors.
For a while we concentrate on the bosonic sector.
As a probe, we utilize D$p$-instanton,
which is a point-like object in Misner space but
$p+1$ dimensional extended object in the extra $d$ directions.%
\footnote{The notation we take is $I=\pm,2,3,\cdots,d+2$
for total space and $\mu = 2,3, \cdots, p+2$ for parallel 
directions to D-instanton. For normal directions, 
we use $A=\pm,a$ with $a = p+3 , \cdots , d+2$.}
As in the previous case, we start from the bulk instanton.
In the covering space we have to introduce infinitely many
image D-instantons, and hence the low energy effective theory
has $\infty \times \infty$ matrices
$X^{A}_{i,j}$, $A^{\mu}_{i,j}$, where $i,j$
are Chan-Paton indices.

When it comes to fractional instanton case,
we rather change the basis of Chan-Paton indices \eqref{basis} as
\begin{align}
X^A_{m,n} &= 
 \sum_{i,j = - \infty}^{\infty} \xi^{im} X^A_{i,j} \xi^{-jn} ~,
&A^{\mu}_{m,n} &= 
 \sum_{i,j = - \infty}^{\infty} \xi^{im} A^{\mu}_{i,j} \xi^{-jn} ~.
\end{align}
The orbifold action acts on these matrices as
\begin{align}
g : X^{\pm}_{i,j} &\to e^{\pm 2 \pi \gamma} X^{\pm}_{i-1,j-1} ~,
&g : A^{\mu}_{i,j} &\to A^{\mu}_{i-1,j-1} ~,
&g : X^{a}_{i,j} &\to X^{a}_{i-1,j-1} ~,
\end{align}
or in the new basis as 
\begin{align}
g : X^{\pm}_{m,n} &\to \xi^{m - n} e^{\pm 2 \pi \gamma} X^{\pm}_{m,n} ~,
&g : A^{\mu}_{m,n} &\to \xi^{m - n} A^{\mu}_{m,n} ~,
&g : X^{a}_{m,n} &\to \xi^{m - n} X^{a}_{m,n} ~.
\end{align}
{}From now on we assign $\xi = e^{2 \pi \gamma}$, then
only $X^{\pm}_{n , n \pm 1}$, $A^{\mu}_{n,n}$, $X^a_{n,n}$
survive the orbifold projection.
This is a similar situation to Euclidean orbifold case
\cite{DM} even though the indices do not obey periodicity condition.
If we use generic $\xi$, then all $X^{\pm}_{m,n}$ are projected
out.\footnote{If we choose $\xi = e^{2 \pi \gamma /N}$
with $N \in {\mathbb Z}$, then $X^{\pm}_{m,m+N}$ would survive.
Since the space-time fermions $S^{\pm}_{m,n}$ in our superstring 
case transform as 
$g:S^{\pm}_{m,n} \to - \xi^{m-n} e^{\pm \pi \gamma} S^{\pm}_{m,n}$,
we may choose $\xi = - e^{\pi \gamma}$ to have 
non-trivial components $S^{\pm}_{m,m \pm 1}$ 
along with $X^{\pm}_{m,m\pm 2}$.}

In general, the effective Lagrangian for the matrices on 
D$p$-instantons is given by
\begin{align}
 {\cal L} = -  \frac14 {\rm Tr} \, 
 \left( F_{\mu \nu} F^{\mu \nu} + 2 D_{\mu} X_A D^{\mu} X^A
    - [ X_A , X_B ] ^2
 \right) ~,
 \label{matrixaction}
\end{align}
where the trace runs over Chan-Paton indices.
For simplicity, we neglect $g$ and $2 \pi \alpha '$ in the Lagrangian.
Now we choose a configuration with every fractional
instantons labeled by $m \in {\mathbb Z}$.
Here we should remark that the Lagrangians with the bases $(i,j)$
and $(m,n)$ might not be the same because the transformation of
basis is not unitary.
In fact, the both Lagrangians would not coincide unless
$\sum _k {\rm Tr} \xi^{kn} = 0$, which is true for 
${\mathbb C}/{\mathbb Z}_N$ case, for instance.%
\footnote{It might be possible that the bulk instanton and
the sum of every fraction instantons are regarded as the same object
if we choose a proper regularization, which we do not know yet.}

Applying the orbifold projection, 
the Lagrangian involving the scalars $X^{\pm}$
is given by
\begin{align}
\begin{aligned}
 {\cal L}_{\pm} = &
   \sum_{n}  \Big[ ( \partial_{\mu} - B_{\mu,n} ) 
     X^+_{n,n+1} ( \partial_{\mu} + B_{\mu,n} )   X^-_{n+1,n}  \\
  & \qquad -   (X^a_{n,n} - X^a_{n+1,n+1})^2  
    | X^+_{n,n+1} |^2 + \frac12 
      ( | X^+_{n,n+1} |^2 - |X^+_{n-1,n}|^2)^2 \Big] ~,
\end{aligned}
\end{align}
where we have defined 
$B_{\mu,n} = i (A_{\mu,n,n} -  A_{\mu,n+1,n+1})$ and 
$|X^+_{n,n+1}|^2 = X^{+}_{n,n+1} X^-_{n+1,n}$.
Due to the orbifold projection, the gauge symmetry left is
\begin{align}
 B_{\mu,n} &\to B_{\mu,n} + \partial_{\mu} \Lambda_n (x^{\mu}) ~,
&X^{+}_{n,n+1} &\to e^{\Lambda_n (x^{\mu})} X^{+}_{n,n+1} ~,
&X^{-}_{n+1,n} &\to e^{-\Lambda_n (x^{\mu})} X^{-}_{n+1,n} ~.
\label{gauge}
\end{align}
We can read off the classical moduli space from 
the saddle point of the potential term as%
\footnote{As mentioned before, we focus on the big bang region
($X^+ , X^- > 0$) of Misner space.}
\begin{align}
X^{\pm}_{n,n\pm 1} &= 0 &{\rm or }& &X^a_{n,n} &= x^a,
  \quad | X^+_{n,n+1} |^2 = \frac{t^2}{2} ~.
\end{align}
The former corresponds to Coulomb branch, where the fractional
instantons are localized at the big bang singularity as expected.
The latter corresponds to Higgs branch, where the center of
the sum of fractional instantons can be placed out of the fixed point.

The situation bears resemblance to the case of ${\mathbb C}/{\mathbb Z}_N$.
There the sum of $N$ different fractional branes can be moved 
away from the fixed point, since the bulk brane is just the sum of them.
Likewise, we can probe the geometry away from the singularity
by using the sum of every fractional instantons, even though
it might differ from the bulk instanton.

This may be understood as follows.
In the boundary state formalism, we can insert a operator into the
boundary state
\begin{align}
  {\rm Tr} {\rm P} \exp \left( - i \int d \sigma M_A \partial_{\tau} X^{A} (\sigma ) \right) ~,
\label{wilson}
\end{align}
where P represents path ordering.
This operator is T-dual to the Wilson line and generates the shift
of the position of corresponding brane.
To preserve the conformal invariance, it is necessary to impose some conditions to $M_A$ .
In the first order of $\alpha '$, the condition that
the beta function vanishes is equal to the equations of
motion to the effective Lagrangian;
$[ M_A , M_B ] = 0$ ($|X^{\pm}_{n,n+1}|^2 = t^2 /2$).
To conclude, even the composite fractional instantons 
cannot be moved by themselves, the center of the sum of them can be shifted
by the insertion of the shift operator \eqref{wilson}.

Let us turn to examine moduli space on the Higgs branch.
On this branch, we can set 
$X^{+}_{n,n+1} = \frac{1}{\sqrt{2}}t e^{ \psi_n}$ and
$X^{-}_{n+1,n} = \frac{1}{\sqrt{2}}t e^{- \psi_n}$.
Making use of the gauge symmetry \eqref{gauge}, we may fix
$\psi_n = \psi$.
This gauge choice leaves unfixed ${\mathbb Z}$ symmetry
as in \cite{APS}, which is generated by
$\exp ( - 2 \pi \gamma \sum_n n Q_n )$.
This means that we have to identify 
$X^{\pm} \sim e^{\pm 2 \pi \gamma} X^{\pm}$, namely
the coordinate $\psi$ has periodicity
$\psi \sim \psi + 2 \pi \gamma$.
In this way, we can identify that the moduli space is 
Misner space as is expected.

The metric of the moduli space is obtained by means of ``kinetic'' term of the Lagrangian. It can be computed as
\begin{align}
 - \sum_{n}  ( \partial_{\mu} - B_{\mu,n} ) 
     X^+_{n,n+1} ( \partial_{\mu} + B_{\mu,n} ) 
     X^-_{n+1,n}   = 
  \sum_i \frac{1}{2} \left[ - (\partial_{\mu} t)^2 + 
 t^2 (\partial_{\mu} \psi_n - B_{\mu , n} )^2 \right] ~.
\end{align}
Decoupling diagonal gauge symmetry forces us to assign $\sum_n B_{\mu , n} = 0$.
This restriction is incorporated into the Lagrangian by 
introducing the Lagrange multiplier 
$\lambda_{\mu}$. Solving the equation of motion for 
$B_{\mu , n}$ and the constraint $\sum_n B_{\mu , n} = 0$, we obtain
\begin{align}
 B_{\mu , n} &= \partial_{\mu} ( \psi_n - \psi ) ~,
 &\sum_n \psi_n = \sum_n (\psi + 2 \pi \gamma n) ~.
\end{align}
In this definition of $\psi$, we have desired periodicity 
$\psi \sim \psi + 2 \pi \gamma$. Then, the kinetic term
becomes
\begin{align}
 \frac{1}{2} \left[ - (\partial_{\mu} t)^2 + 
 t^2 (\partial_{\mu} \psi )^2 \right] ~,
\end{align}
which reproduces the metric in the cosmological region of 
Misner space. 

\subsection{Closed string tachyon condensation}

Taking advantage of the effective theory constructed above,
we are capable of discussing the condensation of closed string
tachyons in the twisted sectors.
Since we are interested in the region near the singularity,
it is enough to think of the deformation to 
leading terms of $X^{\pm}$ as
\begin{align}
 \Delta V = \sum_n m_n^2 |X^+_{n,n+1}|^2  ~.
\end{align}
This term should originate from the three point function
like $\langle \sigma_k \psi^+ \psi^- \rangle$,
where $\sigma_k$ stands for the $k$-th twist vertex operator
of closed string \cite{DM}.
Put it differently, bringing vev to the tachyonic mode causes
the mass deformation.
The three point function would not vanish since the sum of fractional
branes couples to the twisted closed strings as shown before.
It is here of practical advantage to probe the process by fractional instantons,
since this property would be far from obvious in the case of bulk instanton.

For the purpose of the following argument, it helps to re-phrase the mass 
terms as
\begin{align}
 \Delta V = - \sum_n \lambda_n (|X^+_{n,n+1}|^2 - |X^+_{n-1,n}|^2) ~.
 \label{pb}
\end{align}
Then, the saddle point of the potential is shifted as
\begin{align}
 |X^+_{n,n+1}|^2 - |X^+_{n-1,n}|^2 = \lambda_n ~.
 \label{FI}
\end{align}
Hence, $| X^+_{n,n+1}|^2 = t^2_n/2$ depends on $n$ 
after the tachyon condenses.

Let us first examine the simplest case, say,
only $\lambda_1 = \rho^2/2$ is non-zero.
Then, we can just set
\begin{align}
 |X^+_{n,n+1}|^2 & = \frac{t^2}{2}  \quad {\rm for} ~ n \leq 0 ~,
 &|X^+_{n,n+1}|^2 &=  \frac{t^2}{2} + \frac{\rho^2}{2} \quad  {\rm for} ~ n > 0 ~.
\end{align}
The above equations imply that the ungauged 
${\mathbb Z}$ symmetry is reduced to ${\mathbb Z}/2$ symmetry.
Following the previous analysis, we obtain the metric of 
moduli space as
\begin{align}
 &\frac{1}{2} \left[ - A( t ) 
  (\partial_\mu t )^2 + \frac{t^2}{A( t )} (\partial_{\mu} \psi)^2
  \right]~, 
  &A( t ) &=  \frac{1}{2}\left( 1 + \frac{t^2}{t^2 + \rho^2} \right)~.
\end{align}
Notice that the function $A( t )$ connects 1 at $|t|=\infty $ 
and $\frac{1}{2}$ at $|t|=0$. 
In conclusion, Misner space with periodicity $\psi \sim \psi + 2 \pi \gamma$
is now replaced by another Misner space with parameter 
$\psi \sim \psi + 4 \pi \gamma$ near the big bang singularity.
In this way, we can understand that the singularity gets milder by
the tachyon condensation,\footnote{Precisely speaking,
there is still curvature singularity even in Misner space with
larger space cycle. The singularity is resolved if the final
space-time is a part of Minkowski space-time as in the next example.}
or the time period is shortened.%
\footnote{Let us compare two Misner spaces with the periodicity 
$2 \pi \gamma$ and $4 \pi \gamma$ as this example. 
Since the space radius becomes $\gamma L$ when $t=L$ and $t=L/2$, 
respectively,
we interpret this fact as the effective time period is shortened by
the tachyon condensation. This interpretation is consistent with
that in \cite{MS}, where it was claimed that the time-like Liouville 
potential repels the wave function near the big bang singularity.
}
As shown in \cite{APS}, the tip of the cone of
${\mathbb C}/{\mathbb Z}_N$ model is smoothed out by tachyon
condensation, and our case can be regarded as an analogous situation.
Repeating similar deformations, Misner space 
with periodicity $\psi \sim \psi + 2 \pi \gamma$ flows to
Misner space
with periodicity $\psi \sim \psi + 2 \pi \gamma c$ with $c > 1$.

We are in a position to deal with more generic $\lambda_n$.
We can set $|X^+_{0,1}|^2= t^2/2$ to be the smallest one among all
by re-labeling $n$. Defining $\rho_n$ by 
$\rho_{n}^2 - \rho^2_{n-1} = 2 \lambda_n$ with $\rho_0 = 0$,
the solution to \eqref{FI} is given by 
$|X^{\pm}_{n,n+1}|^2 = t_n^2 /2 = (t^2 + \rho_n^2)/2 $.
Due to the generic deformation, the ${\mathbb Z}$ symmetry is completely
broken. Using $N = \sum_n 1$ (which has been neglected elsewhere), 
the kinetic term turns out to be 
\begin{align}
 &\frac{N}{2} \left[ - A( t ) 
  (\partial_\mu t )^2 + \frac{t^2}{A( t )} (\partial_{\mu} \psi)^2
  \right]~, 
  &A( t ) &=  \frac{1}{N} \sum _n \left(\frac{t^2}{t^2 + \rho_n^2} \right)~.
\end{align}
The function $A( t )$ interpolates 1 at $|t|=\infty $ 
and $0$ at $|t|=0$.
Thus the region near the origin of Misner space with 
$2 \pi \gamma$ as the periodicity of $\psi$-cycle is replaced by
Misner space with $\infty$ $\psi$-cycle, namely, 
the future patch of Minkowski space in the Rindler coordinates.
Now that the time-coordinate is shortened by tachyon condensation,
the Minkowski region is not connected to the other regions in
a simple way.

\section{Conclusion} 
\label{conclusion}

We have explored the condensation of 
twisted closed string tachyon in Misner space
accompanied by D-instanton probe.
We spotlight on the big bang region with $t > 0$, that is, the problem involved is whether 
tachyon condensation resolves big bang singularity.
In the bosonic string or superstring theories with
opposite periodic condition of space-time fermion, 
there appears tachyonic modes in the twisted
sectors of closed string.
We have shown that a type of closed string
tachyon is localized in the whisker regions
with the help of classical string trajectories
and Virasoro constraints.
The condensation of the twisted tachyon
affects the cosmological region through the
big bang singularity, which bears a strong resemblance to the localized
tachyon condensation in Euclidean orbifolds \cite{APS}.

In order to see how the small vev of the tachyon changes
the geometry, it is useful to utilize D-instanton probe.
There are ``fractional'' instantons in Misner space, and it is
the sum of them that provides us with a powerful tool to gain further insight.
Two main properties are responsible for this;
(i) D-instanton can couple to closed string tachyon in the twisted sectors, 
and (ii) it is able to be moved away from the big bang singularity.
{}Resorting to the effective action on D-instanton,
it is possible to read off the metric of the background 
both before and after the tachyon condensation.

After tachyon condenses, it has been shown that Misner space with larger space cycle replaces
Misner space with smaller one near the big bang singularity.
In generic tachyon condensation, Misner space is 
replaced by the future patch of Minkowski space in Rindler coordinates.
The tachyon condensation shortens the effective period of time
\cite{MS}, and hence the beginning of the space-time is not directly 
connected with the other regions.

Notice that D-instanton probe stays
reliable merely in the region less than string length \cite{DKPS}.
Therefore, we cannot examine how the tachyon condensation changes
the geometry beyond this region by D-instanton probe.
Far away from the big bang singularity, any geometry change should develop in the gravity regime \cite{APS}.
It may be possible to study this regime by following RG flow or referring to time dynamics.

\subsection*{Acknowledgement}

We would like to thank K.~Hosomichi, K.~Ideguchi, Y.~Matsuo, 
S.~Nagaoka, F.~Yagi and A.~Yamaguchi for useful discussions. 
We also thank the organizers of YITP workshop
``String Theory and Quantum Field Theory'',
where useful discussions were made.


\appendix

\section{Theta and eta functions}
\label{formulae}

We have used the definition of theta functions as
\begin{align}
\begin{aligned}
 \vartheta_1 (\nu|\tau) &= 2 q^{\frac18} \sin \pi \nu
  \prod_{m=1}^{\infty} (1-q^m)(1-zq^m)(1-z^{-1}q^m) ~,
 \\
 \vartheta_2  (\nu|\tau) &= 2 q^{\frac18} \cos \pi \nu
  \prod_{m=1}^{\infty} (1-q^m)(1+zq^m)(1+z^{-1}q^m) ~,
 \\
 \vartheta_3  (\nu|\tau) &=
  \prod_{m=1}^{\infty} (1-q^m)(1+zq^{m-\frac12})(1+z^{-1}q^{m-\frac12}) ~,
 \\
 \vartheta_4  (\nu|\tau) &=
  \prod_{m=1}^{\infty} (1-q^m)(1-zq^{m-\frac12})(1-z^{-1}q^{m-\frac12}) ~,
\end{aligned}
\end{align}
where we have set $q=\exp ( 2 \pi i \tau )$ and $z = \exp (2 \pi i \nu )$.
Their modular transformations are
\begin{align}
\begin{aligned}
 \vartheta_1 \left(\frac{\nu}{\tau} \left| - \frac{1}{\tau} \right.\right)
  &= - i ( - i \tau )^{\frac{1}{2}}
    e^{ \frac{\pi i \nu ^2}{\tau} }
    \vartheta_1 ( \nu | \tau ) ~,
 &\vartheta_2 \left(\frac{\nu}{\tau} \left| - \frac{1}{\tau} \right.\right)
  &= ( - i \tau )^{\frac{1}{2}}
    e^ { \frac{\pi i \nu ^2}{\tau} }
    \vartheta_4 ( \nu | \tau ) ~, \\
 \vartheta_3 \left(\frac{\nu}{\tau} \left| - \frac{1}{\tau} \right.\right)
  &= ( - i \tau )^{\frac{1}{2}}
    e^{ \frac{\pi i \nu ^2}{\tau} }
    \vartheta_3 ( \nu | \tau ) ~,
 &\vartheta_4 \left(\frac{\nu}{\tau} \left| - \frac{1}{\tau} \right.\right)
  &= ( - i \tau )^{\frac{1}{2}}
    e^{ \frac{\pi i \nu ^2}{\tau} }
    \vartheta_2 ( \nu | \tau ) ~.
\end{aligned}
\end{align}
We have also used Dedekind eta function defined as
\begin{align}
 \eta (\tau) &= q^{\frac{1}{24}} \prod_{m=1}^{\infty} (1 - q^m) ~,
 &\eta \left( - \frac{1}{\tau} \right)
 &= ( - i \tau )^{\frac12 } \eta ( \tau ) ~.
\end{align}
\section{D-instantons in superstring theory}

It is also of illustrative interest to construct D-instanton boundary state in superstring cases, 
though the analysis having been made concerns mainly the bosonic part.
In the open string channel the Dirichlet boundary condition is specified as
\begin{align}
\psi^{\pm}  &= - \tilde \psi^{\pm} \quad \text{ at } \sigma = 0 ~,
&\psi^{\pm} &= (-1)^A \tilde \psi^{\pm} \quad \text{ at } \sigma = \pi ~,
\end{align}
where $A=0$ for NS-sector and $A=1$ for R-sector. Through doubling trick, mode expansion is simply 
\begin{align}
 \psi^{\pm} &= \sum_{r}
  b^{\pm}_{-r} e^{-ir(\tau + \sigma)} 
\end{align}
with $r \in {\mathbb Z} + (A+1)/2$.
The oscillators satisfy the anti-commutation relations 
$\{ b^{\pm}_r , b^{\mp}_s \} = - \delta_{r+s}$. 
The orbifold action is defined as
$e^{2 \pi i \gamma \hat J} \psi^{\pm} = e^{\pm 2 \pi \gamma} \psi^{\pm}$, where
\begin{align}
 i J &= - \sum_{r \geq \frac12} b^+_{-r} b^-_r 
       +  \sum_{r \geq \frac12} b^-_{-r} b^+_r
 \end{align}
for NS sector and 
\begin{align}
 i J &= - \sum_{n \geq 1} b^+_{-n} b^-_n 
       +  \sum_{n \geq 0} b^-_{-n} b^+_n
 \end{align}
for R sector.
Equipped with these it is now straightforward to compute the open string one-loop amplitude.%
\footnote{As before, an overall factor $1/N$ with $N=\sum_i 1$ is neglected and also $k=0$ sector is.}
While embedded into the 10d spacetime, including both ghost part 
and GSO projection gives%
\footnote{The factor $(-1)^k$ inside the trace over R-sector comes
from the fact that the space-time fermion has identification
such as $S^{\pm} \sim g^k \cdot S^{\pm}
 = (-1)^k e^{\pm k \pi \gamma} S^{\pm}$ as mentioned before.}
\begin{align}
Z_{mn}(it) &= {\rm Tr}_{{\cal H}_{mn}^{\rm NS}} \sum_k g^k {\textstyle \frac12} (1 + (-1)^F)
  e^{-2\pi t L_0 } - {\rm Tr}_{{\cal H}_{mn}^{\rm R}} \sum_k (-g)^k {\textstyle \frac12} (1 \pm (-1)^F)
  e^{-2\pi t L_0} \nonumber \\
&= \frac{i}{2}\sum^{\infty}_{k = -\infty}\xi^{k({m-n})} 
  2 \sinh (\pi |k| \gamma ) \label{supZ} \\
&\times\frac{\vartheta_4 (i | k |\gamma|it) \vartheta_4 (0|it)^3
             - \vartheta_3(i |k |\gamma|it) \vartheta_3(0|it)^3
             - (-1)^k\vartheta_2 (i |k |\gamma|it) \vartheta_2 ( 0 |it)^3}
             {(8\pi^2 t \alpha')^{\frac{p+1}{2}} 
              \vartheta_1(i |k |\gamma|it) \eta(it)^9} ~, 
              \nonumber
\end{align}
where Neumann boundary condition for
$p+1$ coordinates and Dirichlet boundary condition for $7-p$
coordinates are assumed.
Under the modular transformation $t\rightarrow s=1/t$,
the partition function is changed into
\begin{align}
\begin{aligned}
 &Z_{mn}(is) = \frac{1}{2}\sum^{\infty}_{k = -\infty}\xi^{k({m-n})}2 \sinh (\pi |k| \gamma)\\
 & \quad \times\frac{\vartheta_3 ( | k | \gamma s|is) \vartheta_3 (0|is)^3
             - (-1)^k\vartheta_4 ( | k | \gamma s|is) \vartheta_4 (0|is)^3
             - \vartheta_2 ( | k | \gamma s|is) \vartheta_2 ( 0 |is)^3}
             {(8\pi^2 \alpha')^{\frac{p+1}{2}}s^{\frac{7-p}{2}}
               \vartheta_1( | k | \gamma s|is) \eta(is)^9} ~,
               \label{superD1}
\end{aligned}
\end{align}
which should be reproduced by the overlap between D-instanton boundary states.

Fermionic contribution to the boundary states in $k$-th twisted 
sector of Misner part 
\begin{align}
\begin{aligned}
 ( b^{\pm}_{-r} - i \eta \tilde b^{\pm}_r ) 
| B_{\psi^{\pm}}^f, k, \eta \rangle\rangle_\text P = 0 ~, 
&&\eta = \pm1
\end{aligned}
\end{align}
with $P$ referred to NSNS or RR is solved by
\begin{align}
\begin{aligned}
| B_{\psi^{\pm}}^f, k, \eta \rangle\rangle_{\text{NSNS}}
 &= e^{-i \eta \sum_{n \geq 1} \left(b^+_{-n + 1/2} \tilde b^-_{- n + 1/2}
      + b^-_{-n + 1/2} \tilde b^+_{- n + 1/2} \right) } | 0 \rangle  ~,
\\
| B_{\psi^{\pm}}^f, k, \eta \rangle\rangle_{\text {RR}}
 &= e^{-i \eta \left(\sum_{n \geq 1} b^+_{-n} \tilde b^-_{-n}
      + \sum_{n \geq 0} b^-_{-n} \tilde b^+_{- n} \right) } | 0 \rangle ~.
\end{aligned}
\end{align}
Gathering all components to form the GSO
invariant boundary state, it is thus of the form
\begin{align}
\begin{aligned}
 | B, k \rangle\rangle =& 
  \frac12 \left( | B, k, + \rangle \rangle_{\text {NSNS}}
    - (-1)^k | B, k, - \rangle \rangle_{\text {NSNS}} \right)  \\
    &+ \frac12 \left( | B, k, + \rangle \rangle_{\text {RR}}
    + (-1)^k | B, k, - \rangle \rangle_{\text {RR}} \right) 
\end{aligned}
\end{align}
with
\begin{align}
\begin{aligned}
 | B, k, \eta \rangle\rangle_P = {\cal N}_k | B_{\pm}, k, \eta \rangle\rangle_P \otimes | B_{\text {flat}}, \eta \rangle\rangle_P \otimes
 | B_{\text {gh}}, \eta \rangle\rangle_P ~.
\label{Coeff}
 \end{aligned}
\end{align}
See \cite{HNP2} for the definition of extra direction part.
The overlap of the boundary states is demanded to be equal to 
the open string one-loop amplitude such that
\begin{align}
Z_{mn}  = 
 \frac{\alpha ' \pi}{2} \int_0^{\infty} ds
 \langle \langle m|e^{-\pi{s}(L_0+\tilde{L}_0)}|n\rangle \rangle ~,
\end{align}
where 
\begin{align}
| m \rangle \rangle 
 = \sum_{k=-\infty}^{\infty} B_m^k | B, k \rangle\rangle
\end{align}
is defined.
{}From the above requirement, we obtain
\begin{align}
 B_m^k &= \xi^{- k m} ~,
 &{\cal N}_k &=
 \sqrt{\frac{2 \sinh (\pi |k| \gamma)}{2 \pi ^2 \alpha '}} {\cal N}_p~,
&{\cal N}_p &= \frac{\sqrt{\pi}}{2} (2 \pi \sqrt{\alpha '})^{3 - p} ~,
\end{align}
where ${\cal N}_p$ is the coefficient of the boundary states
for usual D$p$-brane. For bra states, we should use $B_m^{- k}$
instead of $B_m^k$.


\baselineskip=14pt

\end{document}